\documentclass[prd,superscriptaddress,amsfonts,amssymb,amsmath,showpacs,twocolumn]{revtex4-2}
\usepackage{bm}
\usepackage{amsfonts}
\usepackage{latexsym}
\usepackage[latin1]{inputenc}
\usepackage{graphicx}
\usepackage{amsmath}
\usepackage{palatino}
\usepackage{mathpazo}
\usepackage{textcomp}
\usepackage{float}
\usepackage{booktabs}
\usepackage{dcolumn}
\usepackage{hyperref}
\hypersetup{colorlinks,citecolor=blue}
\usepackage{amsmath}
\usepackage{xcolor}
\usepackage{orcidlink}
\usepackage[caption=false]{subfig}
\usepackage{commath}
\def\jnl@style{\it}
\def\aaref@jnl#1{{\jnl@style#1}}

\def\aaref@jnl#1{{\jnl@style#1}}

\def\aj{\aaref@jnl{AJ}}                   % Astronomical Journal
\def\apj{\aaref@jnl{ApJ}}                 % Astrophysical Journal
\def\apjl{\aaref@jnl{ApJ}}                % Astrophysical Journal, Letters
\def\apjs{\aaref@jnl{ApJS}}               % Astrophysical Journal, Supplement
\def\apss{\aaref@jnl{Ap\&SS}}             % Astrophysics and Space Science
\def\aap{\aaref@jnl{A\&A}}                % Astronomy and Astrophysics
\def\aapr{\aaref@jnl{A\&A~Rev.}}          % Astronomy and Astrophysics Reviews
\def\aaps{\aaref@jnl{A\&AS}}              % Astronomy and Astrophysics, Supplement
\def\mnras{\aaref@jnl{Mon.~Not.~Roy.~Astron.~Soc.}}             % Monthly Notices of the RAS
\def\prd{\aaref@jnl{Phys.~Rev.~D}}        % Physical Review D
\def\prc{\aaref@jnl{Phys.~Rev.~C}}  % Physical Review C
\def\prl{\aaref@jnl{Phys.~Rev.~Lett.}}    % Physical Review Letters
\def\qjras{\aaref@jnl{QJRAS}}             % Quarterly Journal of the RAS
\def\skytel{\aaref@jnl{S\&T}}             % Sky and Telescope
\def\ssr{\aaref@jnl{Space~Sci.~Rev.}}     % Space Science Reviews
\def\zap{\aaref@jnl{ZAp}}                 % Zeitschrift fuer Astrophysik
\def\nat{\aaref@jnl{Nature}}              % Nature
\def\aplett{\aaref@jnl{Astrophys.~Lett.}} % Astrophysics Letters
\def\apspr{\aaref@jnl{Astrophys.~Space~Phys.~Res.}} % Astrophysics Space Physics Research
\def\physrep{\aaref@jnl{Phys.~Rep.}}      % Physics Reports
\def\physscr{\aaref@jnl{Phys.~Scr}}       % Physica Scripta
\def\commat{\aaref@jnl{Comm.~Math.~Phys.}}              % Communications in Mathematical Physics
\def\science{\aaref@jnl{Science}}               % Science
\def\cqg{\aaref@jnl{Classical Quant.~Grav.}}            % Classical and Quantum Gravity
\def\jpcs{\aaref@jnl{JPCS}}                                     % Journal of Physics Conference Series
\def\ijmpd{\aaref@jnl{Int.~J.~Mod.~Phys.~D}}                    % International Journal of Modern Physics D
\def\grg{\aaref@jnl{Gen.~Relat.~Gravit.}}               % General Relativity and Gravitation
\def\rpp{\aaref@jnl{Rep.~Prog.~Phys.}}          % Reports on Progress in Physics
\def\npa{\aaref@jnl{Nucl.~Phys.~A}}        % Nuclear Physics A
\def\lrr{\aaref@jnl{Living Rev.~Rel.}}                   % Living reviews in relativity
\def\jcap{\aaref@jnl{J.~Cosmology Astropart.~Phys.}}    % Journal of cosmology and astroparticle physics
\def\rmp{\aaref@jnl{Rev.~Mod.~Phys.}}   %Reviews of modern physics
\def\epjc{\aaref@jnl{Eur.~Phys.~J.~C}}

\allowdisplaybreaks[1]
\begin{document}

\color{black}       %% For one column

\title{Cosmic jerk parameter in symmetric teleparallel cosmology}
%\end{document}
\author{M. Koussour\orcidlink{0000-0002-4188-0572}}
\email{pr.mouhssine@gmail.com}
\affiliation{Quantum Physics and Magnetism Team, LPMC, Faculty of Science Ben
M'sik,\\
Casablanca Hassan II University,
Morocco.}

\author{S. Dahmani\orcidlink{0000-0001-6247-4496}}
\email{dahmani.safae.1026@gmail.com}
\affiliation{Laboratory of physics of matter and radiations,\\
University Mohammed First, Oujda, Morocco.}

\author{M. Bennai\orcidlink{0000-0003-1424-7699}}
\email{mdbennai@yahoo.fr }
\affiliation{Quantum Physics and Magnetism Team, LPMC, Faculty of Science Ben
M'sik,\\
Casablanca Hassan II University,
Morocco.} 
\affiliation{Lab of High Energy Physics, Modeling and Simulations, Faculty of
Science,\\
University Mohammed V-Agdal, Rabat, Morocco.}

\author{T. Ouali\orcidlink{0000-0002-0776-1625}}
\email{ouali1962@gmail.com}
\affiliation{Laboratory of physics of matter and radiations,\\
University Mohammed First, Oujda, Morocco.}
%
%%%%%%%%%%%%%%%%%%%%%%%%%%%%%%%%%%%%%  DATE  %%%%%%%%%%%%%%%%%%%%%%%%%%%%%%%%%%%%
\date{\today}
\begin{abstract}
In this paper, we have examined the recently proposed modified symmetric
teleparallel gravity, in which gravitational Lagrangian is given by an
arbitrary function of non-metricity scalar $Q$. We have considered a constant jerk parameter to express the Hubble rate. Moreover, we have
used 31 points of OHD datasets and 1701 points of Pantheon+
datasets to constraint our model parameters by means of the Markov Chain
Monte Carlo analysis. The mean values and the best fit obtained give a
consistent Hubble rate and deceleration parameter compared to the
observation values. In order to study the current accelerated expansion
scenario of the Universe with the presence of the cosmological fluid as a
perfect fluid, we have considered two forms of teleparallel gravity. We have
studied the obtained field equations with the proposed forms of $f(Q)$
models, specifically, linear $f\left( Q\right) =\alpha Q+\beta $ and
non-linear $f\left( Q\right) =Q+mQ^{n}$ models. Next, we have discussed the
physical behavior of cosmological parameters such as energy density,
pressure, EoS parameter, and deceleration parameter for both model. To
ensure the validity of our proposed cosmological models, we have checked all
energy conditions. The properties of these parameters confirm that our
models describe the current acceleration of the expansion of the Universe.
This result is also corroborated by the energy conditions criteria. the
Finally, the EoS parameter for both models indicates that the cosmological
fluid behaves like a quintessence dark energy model.
\end{abstract}

\maketitle

\date{\today}
\section{Introduction}

\label{sec1}

A set of recent observations of Type Ia Supernova (SN Ia) \cite{SN1, SN2},
Cosmic Microwave Background (CMB) \cite{CMB1, CMB2}, large scale structure 
\cite{LS1, LS2}, Baryonic Acoustic Oscillations (BAO) \cite{BAO1, BAO2}, and
Wilkinson Microwave Anisotropy Probe (WMAP) experiment \cite{WMAP1, WMAP2}
show an unexpected behavior of cosmic expansion. The cause of this late-time
cosmic acceleration is one of the most significant unresolved problems in
science today. The introduction of a new type of energy known as dark energy
(DE), which makes up a substantial portion of the Universe's total energy,
is the first theory puted up to explain the enigma of the cosmic acceleration
within the context of GR. DE can result a negative pressure ($p<0)$ or
equivalently its EoS (Equation of State) parameter is negative ($\omega
\equiv \frac{p}{\rho }<0$), where $\rho $ is the energy density of the
Universe. According to recent WMAP9 \cite{WMAP9} observations, collecting
data from $H_{0}$ measurements, SN Ia, CMB, and BAO, prove that the present
value of the EoS parameter is $\omega _{0}=-1.084\pm 0.063$. Also, in the
year 2018 Planck collaboration suggests that $\omega
_{0}=-1.028\pm 0.032$ \cite{Planck2018}. The cosmological constant ($\Lambda 
$) with $\omega _{\Lambda }=-1$ that Einstein incorporated into the field
equations in another context provides the best description for DE \cite%
{Weinberg}. Other DE alternatives, such as quintessence DE models with EoS
parameter value in the range $-1<\omega <-\frac{1}{3}$ \cite{quintessence}
and phantom energy models with EoS parameter value $\omega <-1$ \cite%
{phantom}, have been developed in response to the issues related to its
expected order of magnitude from quantum gravity contrasted to the observed
value. Always within the framework of GR, and motivated by these models
other more attractive and interesting dynamical DE models have been
suggested \cite{kessence, Chameleon, tachyon, Cgas1, Cgas2, ouali2015,bouhmadi2017,bouhmadi2018,ouali2019,ouali2021}. The most widely accepted cosmological model today
is the "standard model" or "$\Lambda $CDM" ($\Lambda $+ Cold Dark Matter),
which states that at the beginning of the Universe, photons predominated and
that this period is known as the "radiation-dominated era". With cosmic
expansion, however, another period known as the "matter-dominated era"
appeared while $\Lambda $ was initially slow. However, as the Universe began
to get older and expand faster between five and six billion years ago, $%
\Lambda $ eventually predominated matter \cite{Capozziello}. A cosmic jerk
is probably what caused the Universe to transition from its early
decelerating phase to its current accelerating phase. For several models
with a positive sign~for the jerk parameter and a negative sign for the
deceleration parameter, this transition happens in the Universe \cite%
{Blandford, Chiba, Sahin, Visser1, Visser2}. Jerk parameters are an important tool to distinguish between dynamical models even if they are equivalent kinematically. In the case of $\Lambda$CDM, the matter-dominated phase is an attractor in the past while the density of dark energy is an attractor in the future. This behavior is in general not true for other alternative models, such as modified gravity, which are kinematically degenerate to $\Lambda$CDM. The degeneracy of the jerk parameter comes from its definition as a third order differential of the scale factor. This definition gives a large choice of solutions. This is the kinematic degeneracy between different dynamical models. For the flat FLRW metric,
the cosmographic jerk parameter is given as,%
\begin{equation}
j=\frac{{\dddot{a}}}{aH^{3}}=1.  \label{jerk}
\end{equation}%
where the Hubble parameter, $H=\frac{{\dot{a}}}{a}$, represents the
expansion rate of the Universe. Several authors have suggested applications
of the jerk parameter as a means of reconstructing cosmological models in
various cosmological contexts \cite{Alam, Rapetti}. The cosmological implications of $f(T, B)$ gravity have been investigated using the jerk parameter in Ref. \cite{Zubair}. Some kinematic models have been constrained by the most current observational data by using the jerk parameter \cite{Lu}.

In the space-time manifold, we can outline the gravitational interactions
using three types of concepts namely curvature $R$, torsion $T$, and
non-metricity $Q$. In the famous theory of GR, the concept responsible for
gravitational interactions is the curvature of space-time. The teleparallel
and symmetric teleparallel equivalents of GR are two additional options
that, together with torsion and non-metricity, provide an analogous
explanation of GR. It can be said that the $f(R)$ gravity mentioned above is
a modification of curvature-founded gravity (GR) with zero torsion and
non-metricity. Also, we can say that
the $f(Q)$ gravity is a modification of the Symmetric Teleparallel
Equivalent of GR (STEGR) with zero torsion and curvature, in other words, $%
f(Q)$\ gravity is equivalent to GR in flat space \cite{Xu} (for more
details, see Sec. \ref{sec2}). Numerous articles on $f(Q)$ gravity have been
written and published. The energy conditions and cosmography under $f(Q)$
gravity have been tested by Mandal et al. \cite{Mandal1, Mandal2}. Harko et
al. studied the coupling matter in modified $Q$ gravity by presuming a
power-law function \cite{Harko1}. Dimakis et al. examined quantum cosmology
for a $f(Q)$ polynomial model \cite{Dimakis}. Also, the idea of holographic dark energy in $f(Q)$ symmetric teleparallel gravity has been discussed by Shekh \cite{Shekh}, while the late-time acceleration of the Universe can be described by $f(Q)$ gravity using the hybrid expansion law \cite{Koussour1}, and the anisotropic nature of space-time in $f(Q)$ gravity has been investigated in \cite{Koussour2}.

Our aim in this paper is to exploit the above cosmographic jerk parameter
relation for the flat FLRW metric to construct cosmological models
as alternatives to the cosmic acceleration in the framework of $f(Q)$
gravity (or the so-called symmetric teleparallel gravity) that has been
recently proposed \cite{Jimenez1, Jimenez2} in which the non-metricity
scalar $Q$\ characterize the gravitational interactions. The acceleration of
the expansion of the Universe may also be described by the parametrization method in
the context of standard cosmology or in modified gravity. This approach is commonly referred to as the model-independent way study of cosmological models \cite{para1, para2}. The parameterization method has no effect on the theoretical framework for this study and clearly provides solutions to the field equation. It also has the benefit of reconstructing the cosmic evolution of the Universe and explaining some of its phenomena \cite{para3}. In the current
paper, we consider this kind of parametrization in the case of $f(Q)$
modified gravity, especially, the parameterization of the jerk parameter in Eq. (\ref{jerk}), by deriving the general solution and adding the extra constraint to solve the field equations in $f(Q)$ gravity. The late-time behaviors of such modified gravity, which are kinematically degenerate to $\Lambda$CDM, and its consistency with the current acceleration of the Universe are the main motivation of our paper. From this perspective, we consider two forms of $f(Q)$ gravity, specifically, linear $f\left( Q\right) =\alpha Q+\beta $ and
non-linear $f\left( Q\right) =Q+mQ^{n}$ models, where $\alpha$, $\beta$, $m$, and $n$ are free parameters.

The current document is set up as follows: In Sec. \ref{sec2} we discuss
some basics of $f(Q)$ gravity and derive the field equations for the
cosmological fluid. Cosmological solutions of the field equations are
described with the help of the jerk model parameter in Sec. %
\ref{sec3}. In the same section, we use OHD and Pantheon+ datasets to
estimate the model parameters by means of the Markov Chain Monte Carlo and
we discussed the energy conditions for $f(Q)$ gravity in the latter.
Further, in Sec. \ref{sec4} we discuss different behaviors of our
cosmological models according to the choice of linear and non linear form of 
$f(Q)$\ function. Finally, Sec. \ref{sec5} is used to recapitulate and
conclude the results.

\section{Some basics of $f(Q)$ gravity}

\label{sec2}

A general affine connection $\bar{\Gamma}$ in differential geometry can be
decomposed into the following three independent components: the Christoffel
symbol ${\Gamma ^{\gamma }}_{\mu \nu }$, the contortion tensor ${C^{\gamma }}%
_{\mu \nu }$ and the disformation tensor ${L^{\gamma }}_{\mu \nu }$ and is
given by \cite{Xu} 
\begin{equation}
\bar{\Gamma}{^{\gamma }}_{\mu \nu }={\Gamma ^{\gamma }}_{\mu \nu }+{%
C^{\gamma }}_{\mu \nu }+{L^{\gamma }}_{\mu \nu },  \label{WC}
\end{equation}%
where ${\Gamma ^{\gamma }}_{\mu \nu }\equiv \frac{1}{2}g^{\gamma \sigma
}\left( \partial _{\mu }g_{\sigma \nu }+\partial _{\nu }g_{\sigma \mu
}-\partial _{\sigma }g_{\mu \nu }\right) $ is the Levi-Civita connection of
the metric $g_{\mu \nu }$, the contorsion tensor ${C^{\gamma }}_{\mu \nu }$
can be written as ${C^{\gamma }}_{\mu \nu }\equiv \frac{1}{2}{T^{\gamma }}%
_{\mu \nu }+T_{(\mu }{}^{\gamma }{}_{\nu )}$, with the torsion tensor
described as ${T^{\gamma }}_{\mu \nu }\equiv 2\Sigma {^{\gamma }}_{[\mu \nu
]}$. Finally, the non-metricity tensor $Q_{\gamma \mu \nu }$ is used to
obtain the disformation tensor ${L^{\gamma }}_{\mu \nu }$ as, 
\begin{equation}
{L^{\gamma }}_{\mu \nu }\equiv \frac{1}{2}g^{\gamma \sigma }\left( Q_{\nu
\mu \sigma }+Q_{\mu \nu \sigma }-Q_{\gamma \mu \nu }\right) .  \label{L}
\end{equation}

The non-metricity tensor $Q_{\gamma \mu \nu }$ can be written
as, 
\begin{equation}
Q_{\gamma \mu \nu }=\nabla _{\gamma }g_{\mu \nu }\,,
\end{equation}%
\begin{equation}
Q_{\gamma }={{Q_{\gamma }}^{\mu }}_{\mu }\,,\qquad \widetilde{Q}_{\gamma }={%
Q^{\mu }}_{\gamma \mu }\,.
\end{equation}

It is also useful to introduce the superpotential tensor (non-metricity
conjugate) as, 
\begin{equation}
4{P^{\gamma }}_{\mu \nu }=-{Q^{\gamma }}_{\mu \nu }+2Q{_{(\mu \;\;\nu
)}^{\;\;\;\gamma }}+Q^{\gamma }g_{\mu \nu }-\widetilde{Q}^{\gamma }g_{\mu
\nu }-\delta _{\;(\mu }^{\gamma }Q_{\nu )}\,,
\end{equation}%
where the non-metricity scalar can be obtained as, 
\begin{equation}
Q=-Q_{\gamma \mu \nu }P^{\gamma \mu \nu }\,.
\end{equation}

 Within the current context, the connection is assumed to be torsion- and
curvature-free, making the contorsion tensor ${C^{\gamma }}_{\mu \nu }=0$.

The modified Einstein-Hilbert action in symmetric teleparallel gravity can
be considered as \cite{Jimenez1, Jimenez2} 
\begin{equation}
S=\int\sqrt{-g}d^{4}x \left[ \frac{1}{2}f(Q)+\mathcal{L}_{m}\right],
\label{action}
\end{equation}%
where $f(Q)$ can be expressed as a arbitrary function of non-metricity
scalar $Q$, $g$ is the determinant of the metric tensor $g_{\mu \nu }$ and $%
\mathcal{L}_{m}$ is the matter Lagrangian density assumed to be only
dependent on the metric and independent of the affine connection.

The description of the matter energy-momentum tensor is 
\begin{equation}
T_{\mu \nu }=-\frac{2}{\sqrt{-g}}\frac{\delta (\sqrt{-g}\mathcal{L}_{m})}{%
\delta g^{\mu \nu }}\,.
\end{equation}

The gravitational field equations are now found by varying the modified
Einstein-Hilbert action (\ref{action}) with regard to the metric tensor $%
g_{\mu \nu }$ 
\begin{widetext}
\begin{equation}
\frac{2}{\sqrt{-g}}\nabla _{\gamma }(\sqrt{-g}f_{Q}P^{\gamma }{}_{\mu \nu })+%
\frac{1}{2}fg_{\mu \nu }+f_{Q}(P_{\nu \rho \sigma }Q_{\mu }{}^{\rho \sigma
}-2P_{\rho \sigma \mu }Q^{\rho \sigma }{}_{\nu })=-T_{\mu \nu }.  \label{F}
\end{equation}%
\end{widetext}where we have used the following notation $f_{Q}={df}/{dQ}$. 
Thus, when there is no hypermomentum, the connection field equations
read,
\begin{equation}
\nabla ^{\mu }\nabla ^{\nu }\left( \sqrt{-g}\,f_{Q}\,P^{\gamma }\;_{\mu \nu
}\right) =0.  \label{fe2}
\end{equation}

According to the cosmological principle, our Universe is homogeneous and
isotropic at a large scale. Here, we consider the standard
Friedmann-Lemaitre-Robertson-Walker (FLRW) metric, 
\begin{equation}
ds^{2}=-dt^{2}+a^{2}(t)\left[ dx^{2}+dy^{2}+dz^{2}\right] ,  \label{FLRW}
\end{equation}%
where $a(t)$ is the scale factor which measures how the distance between two
objects varies with time in the expanding Universe. The non-metricity scalar 
$Q$ is given by 
\begin{equation}
Q=6H^{2}.
\end{equation}

In cosmology, the contents of the Universe are often considered to be filled
with a perfect fluid, i.e. a fluid without viscosity. In this case, the
stress-energy momentum tensor of the cosmological fluid is given by 
\begin{equation}
T_{\mu \nu }=(p+\rho )u_{\mu }u_{\nu }+pg_{\mu \nu },  \label{PF}
\end{equation}%
where $p\ $symbolizes the isotropic pressure with cosmological fluid and $%
\rho $ symbolizes the energy density of the Universe. Here, $u^{\mu }=\left(
1,0,0,0\right) $ represent the components of the four velocities of the
cosmological fluid.

To derive the modified Friedmann equations in $f(Q)$ gravity in the case of
a Universe described by the FLRW metric (\ref{FLRW}) in the
coincident gauge i.e. ${\Gamma ^{\gamma }}_{\mu \nu }=-{L^{\gamma }}_{\mu
\nu }$\textbf{\ }\cite{Jimenez1, Jimenez2}, we use Eqs. (\ref{F}) and (\ref%
{PF}) to obtain, 
\begin{equation}
3H^{2}=\frac{1}{2f_{Q}}\left( -\rho +\frac{f}{2}\right) ,  \label{F1}
\end{equation}%
\begin{equation}
\dot{H}+3H^{2}+\frac{\dot{f}_{Q}}{f_{Q}}H=\frac{1}{2f_{Q}}\left( p+\frac{f}{2%
}\right) ,  \label{F2}
\end{equation}%
where the dot $(\overset{.}{})$ denotes the derivative with respect to the
cosmic time $t$. The energy conservation equation of the stress-energy momentum tensor writes
\begin{equation}
\dot{\rho}+3H(\rho +p)=0.
\end{equation}

Now, by eliminating the term $3H^{2}$\ from the previous two Eqs. (\ref{F1})
and (\ref{F2}), we get the following evolution equation for $H$, 
\begin{equation}
\dot{H}+\frac{\dot{f}_{Q}}{f_{Q}}H=\frac{1}{2f_{Q}}\left( p+\rho \right) .
\label{evo}
\end{equation}

Again, by using Eqs. (\ref{F1}) and (\ref{F2}), we obtain the expressions of
the energy density $\rho $ and the isotropic pressure $p$, respectively as, 
\begin{equation}
\rho =\frac{f}{2}-6H^{2}f_{Q},
\label{F11}
\end{equation}%
\begin{equation}
p=\left( \dot{H}+3H^{2}+\frac{\dot{f_{Q}}}{f_{Q}}H\right) 2f_{Q}-\frac{f}{2}.
\label{F22}
\end{equation}

Using Eqs. (\ref{F2}) and (\ref{evo}) we can rewrite the cosmological
equations similar to the standard Friedmann equations in GR, by adding the
concept of an effective energy density $\rho _{eff}$ and an effective
isotropic pressure $p_{eff}$ as, 
\begin{equation}
3H^{2}=\rho _{eff}\,=-\frac{1}{2f_{Q}}\left( \rho -\frac{f}{2}\right) ,
\label{eff1}
\end{equation}%
\begin{equation}
2\dot{H}+3H^{2}=-p_{eff}\,=-\frac{2\dot{f_{Q}}}{f_{Q}}H+\frac{1}{2f_{Q}}%
\left( \rho +2p+\frac{f}{2}\right) .  \label{eff2}
\end{equation}

The equations above can be interpreted as an additional component of a
modified energy-momentum tensor $T_{\,\mu \nu }^{eff}$ due to the
non-metricity terms that behaves as an effective dark energy fluid.
Furthermore, the gravitational action (\ref{action}) is reduced to the
standard Hilbert-Einstein form in the limiting case $f\left( Q\right) =-Q$.
For this choice, we regain the so-called STEGR \cite{Lazkoz}, and Eqs. (\ref%
{eff1}) and (\ref{eff2}) reduce to the standard Friedmann equations of GR, $%
3H^{2}=\rho $, and $2\dot{H}+3H^{2}=-p$, respectively.

\section{Cosmological solutions and energy conditions}

\label{sec3}

The equations (\ref{F1})-(\ref{F2}) are a system of two independent
equations with four unknowns, namely: $\rho $, $p$, $f$, and $H$. It is
therefore difficult to solve it completely without adding other equations or
constraints to the model. Here we will build our model using the parameterization of the jerk parameter. Also in the literature, an additional approach to define degenerate
models to $\Lambda $CDM is jerk parameter. Thus, the
general solution of Eq. (\ref{jerk}) can be expressed as \cite{Zubair,
Chakrabarti}, 
\begin{equation}
a\left( t\right) =\left( Ce^{\lambda t}+De^{-\lambda t}\right) ^{\frac{2}{3}%
},  \label{a}
\end{equation}%
which is an increasing function of the cosmic time describing the accelerated
behavior of the Universe. In Eq. (\ref{a}), $C$, $D$, and $\lambda $ are
model parameters that will be constrained by observational data. To obtain
cosmological results that give a direct comparison of model predictions with
observational data, also, one of the most useful approaches in cosmology is
to determine cosmological parameters in terms of cosmological redshift $z$
in lieu of the cosmic time $t$. For this, we use the relation between the
scale factor of the Universe and the cosmological redshift as $a\left(
t\right) =\left( 1+z\right) ^{-1}$, where the value of the scale factor at
present is $a\left( 0\right) =1$. However and in order to illustrate our
reconstruction $f(Q)$ gravity from the jerk parameter, we assume $C=-D$.

From Eq. (\ref{a}) we obtain the cosmic time in terms of the cosmological
redshift as%
\begin{equation}
t\left( z\right) =\frac{1}{\lambda }\sinh ^{-1}\left( \frac{(1+z)^{-3/2}}{2C}%
\right).  \label{t}
\end{equation}

Now, using Eq. (\ref{a}), the Hubble parameter takes the form 
\begin{equation}
H\left( t\right) =\frac{2\lambda \left( e^{2\lambda t}+1\right) }{3\left(
e^{2\lambda t}-1\right) }.  \label{H}
\end{equation}

Thus, from Eqs. (\ref{t}) and (\ref{H}) we obtain the Hubble parameter in
terms of the cosmological redshift as 
\begin{equation}
H\left( z\right) =\frac{2\lambda \left( e^{2\sinh ^{-1}\left( \frac{1}{%
2C(z+1)^{3/2}}\right) }+1\right) }{3\left( e^{2\sinh ^{-1}\left( \frac{1}{%
2C(z+1)^{3/2}}\right) }-1\right) }.  \label{Hz}
\end{equation}

The deceleration parameter $q$ that describes the evolution of the Universe
can be gained as a function of the cosmological redshift $z$ as \cite{Xu} 
\begin{equation}
q\left( z\right) =-1+\left( 1+z\right) \frac{1}{H\left( z\right) }\frac{%
dH\left( z\right) }{dz}.  \label{q}
\end{equation}

Using Eqs. (\ref{Hz}) and (\ref{q}), we get 
\begin{equation}
q\left( z\right) =-\frac{-4e^{2\sinh ^{-1}\left( \frac{1}{2C(z+1)^{3/2}}%
\right) }+e^{4\sinh ^{-1}\left( \frac{1}{2C(z+1)^{3/2}}\right) }+1}{\left(
e^{2\sinh ^{-1}\left( \frac{1}{2C(z+1)^{3/2}}\right) }+1\right) ^{2}}
\label{qz}
\end{equation}

Now, from expressions of the Hubble parameter and the deceleration parameter
in terms of the cosmological redshift given by Eqs. (\ref{Hz}) and (\ref{qz}%
), we will concentrate in the next section on the constraint of the model
parameters using observational data.

\subsection{Observational constraints}
To constrain the jerk model with two free parameters i.e. C and $\lambda$, we perform the minimization of chi-square function, $\chi ^{2}$, using the Markov Chain Monte Carlo (MCMC) algorithm \cite{MCMC}, where $\chi^{2}=-2\ln {(\mathcal{L}_{max})}$ with $\mathcal{L}$ is the likelihood function. We use the following datasets:

\textbf{Observational Hubble datasets (OHD)}: is a typical compilation of $31$ measures from Hubble data collected using the differential age approach (DA). The expansion rate of the Universe at redshift z may be calculated using this approach \cite{Sharov, data2, data3}. The chi-square of OHD datasets is given by 
\begin{equation}
\chi _{OHD}^{2}(C,\lambda )=\sum_{i=1}^{31}\dfrac{\left[
H_{obs}(z_{i})-H_{th}(C,\lambda ,z_{i})\right] ^{2}}{\sigma ^{2}(z_{i})},
\end{equation}%
where $H_{th}(C,\lambda ,z_{i})$, $H_{obs}(z_{i})$ and $\sigma (z_{i})$ are the theoretical values, the observed values predicted by the model of the Hubble parameter at redshift $z_{i}$ and the standard deviation, respectively. $C$ and $\lambda $ are the free parameters of our theoretical model.\\

\textbf{Pantheon+}: we use the Pantheon+ datasets with 1701 light curves from 1550 Supernovae Type Ia (denoted as SNIa) in the redshift range $0.0001\leqslant z\leqslant 2.26$ \cite{data4}, with the theoretical distance modulus 
\begin{equation}
\mu _{th}(z)=5log_{10}D_{L}(z)+25,
\end{equation}
$D_{L}$ is the luminosity distance defined by 
\begin{equation}
D_{L}(z)=(1+z)\int_{0}^{z}\frac{cdz^{\prime }}{H(z^{\prime })},
\end{equation}%
where $c$ is the speed of light

The $\chi^2$ of Pantheon+ is given by

\begin{widetext}
\begin{equation}
\chi _{SNe}^{2}(C,\lambda,z )=(\mu _{th}(C,\lambda ,z)-\mu
_{obs}(z))\mathcal{C}^{-1}(\mu _{th}(C,\lambda ,z)-\mu
_{obs}(z))^T,
\label{4i}
\end{equation}%
\end{widetext}

where $\mu _{obs}(z)$ is the observational distance modulus, with $\mu _{obs}(z)= m_{obs}-M$, $m_{obs}$ and M are the observed apparent magnitude and the absolute magnitude, respectively. $\mu_{th}(C,\lambda ,z)$ is the theoretical distance modulus value and $\mathcal{C}$ is the covariance matrix.

The $\chi _{tot}^{2}$ will be the sum  of $\chi ^{2}$ of the OHD datasets, $\chi _{OHD}^{2}$, and of the Pantheon+ datasets,  $\chi _{SNIa}^{2}$

\begin{equation}
\chi _{tot}^{2}=\chi _{OHD}^{2}+\chi _{SNIa}^{2}.
\end{equation}

 Fig. \ref{H+SNe} displays the 1D posterior distributions and the 2D confidence contours at 1$\sigma$, 2$\sigma$ and 3$\sigma$ for our model parameters i.e. $C$ and $\lambda $ using the OHD + Pantheon+ datasets, respectively. Tab. \ref{tab} shows the mean$\pm 1\sigma$ values for our model and $\Lambda$CDM parameters, by using the data combination OHD + Pantheon+. We obtain  $h=0.71\pm 0.0082$, $\Omega_m=0.282\pm 0.0118$ and  $C=0.3133\pm0.0095$, $\lambda=90.74\pm 1.632$ for $\Lambda$CDM and jerk model, respectively. Moreover, we notice that the value of the absolute magnitude, $M$, for our model i.e. $M=-19.324\pm 0.0238$ is consistent with the value obtained by the standard model i.e. $M=-19.3235\pm0.02317$.

In Fig. \ref{Hubblecond} we show the evolution of H(z) for the jerk model compared to the standard model, $\Lambda$CDM, using the results obtained by OHD + Pantheon+ datasets (see Tab. \ref{tab} ).

Fig. \ref{Muzcond} shows the evolution of the theoretical distance modulus $\mu(z)$ for our model compared to the $\Lambda $CDM model. Using the best-fit cosmological parameters obtained by OHD + Pantheon+ datasets (see Tab. \ref{tab}).
Fig. \ref{fig2} displays the evolution of the deceleration parameter we use the results obtained by OHD + Pantheon+ datasets. It seems obvious that our Universe recently underwent a transition from a decelerated phase to an accelerated phase. According to the model parameter values constrained by the OHD + Pantheon+ datasets, the transition redshift is $z_{tr}=0.7206$. In addition, the current value of the deceleration parameter is $q_{0}=-0.5746$.

\begin{widetext}

\begin{figure}[h]
\centerline{\includegraphics[scale=0.70]{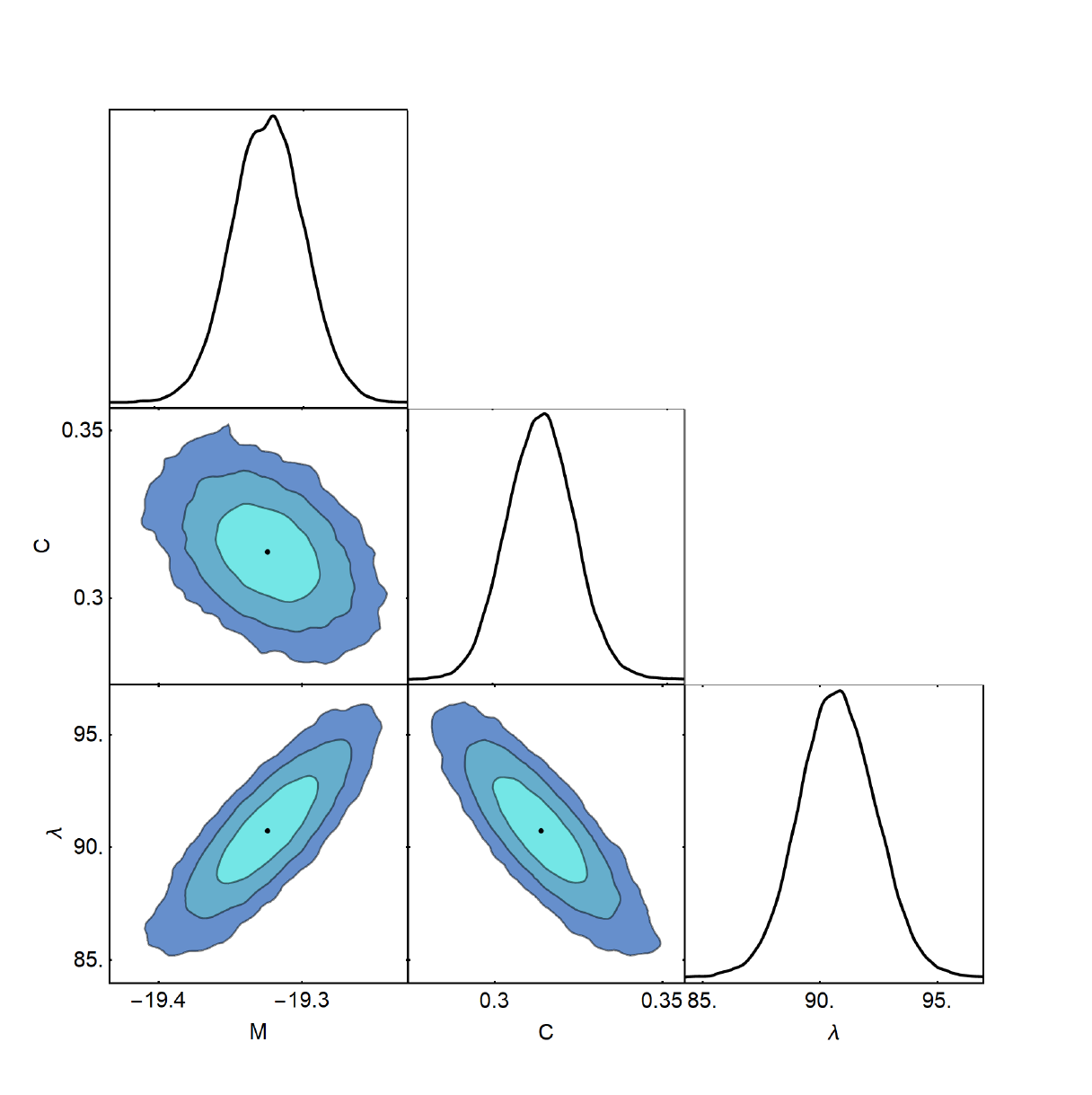}}
\caption{The posterior distributions at 1$\sigma$, 2$\sigma$ and 3$\sigma$ for the jerk model using OHD+Pantheon+ datasets. }
\label{H+SNe}
\end{figure}

\begin{figure}[h]
\centerline{\includegraphics[scale=0.60]{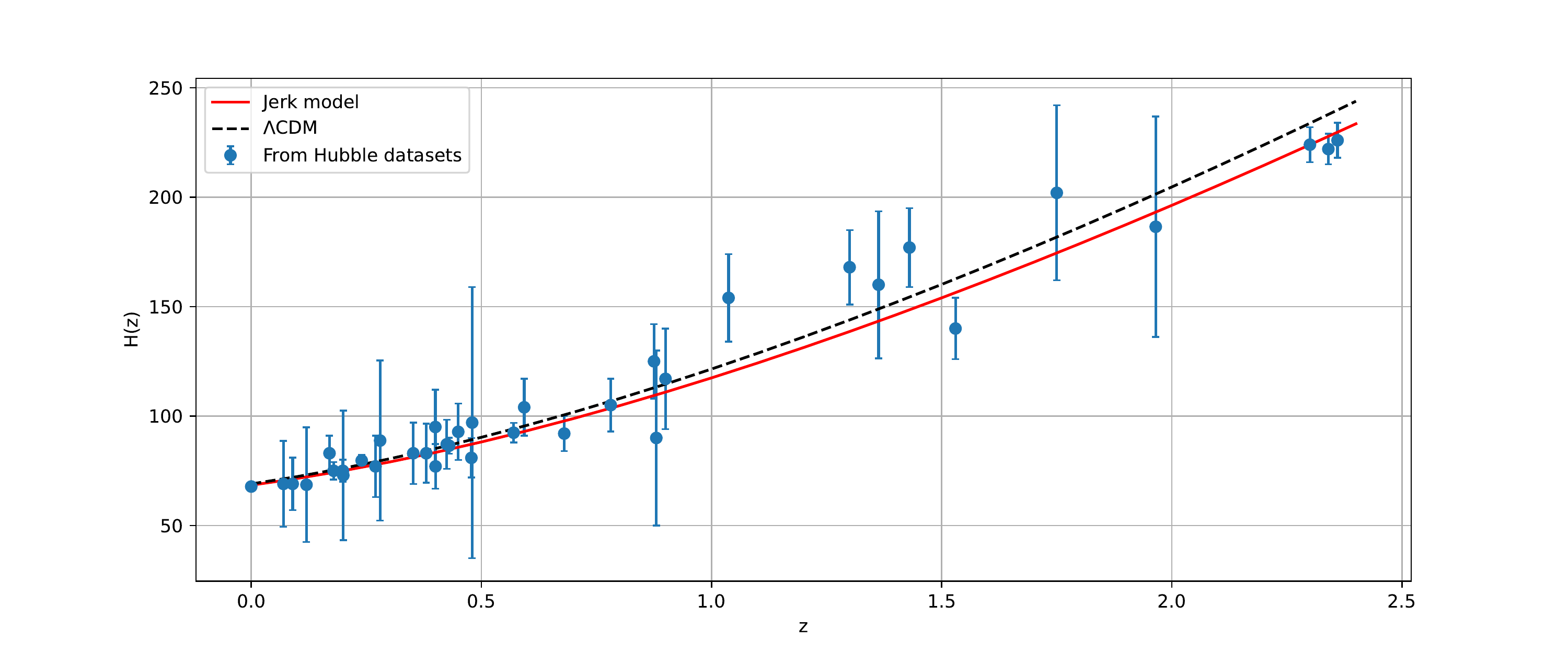}}
\caption{A good fit to the 31 points of the OHD datasets is displayed in the plot of $H(z)$ versus the redshift $z$ for our jerk model, which is shown in red, and $\Lambda$CDM, which is shown in black dashed lines.}
\label{Hubblecond}
\end{figure}

\begin{figure}[h]
\centerline{\includegraphics[scale=0.60]{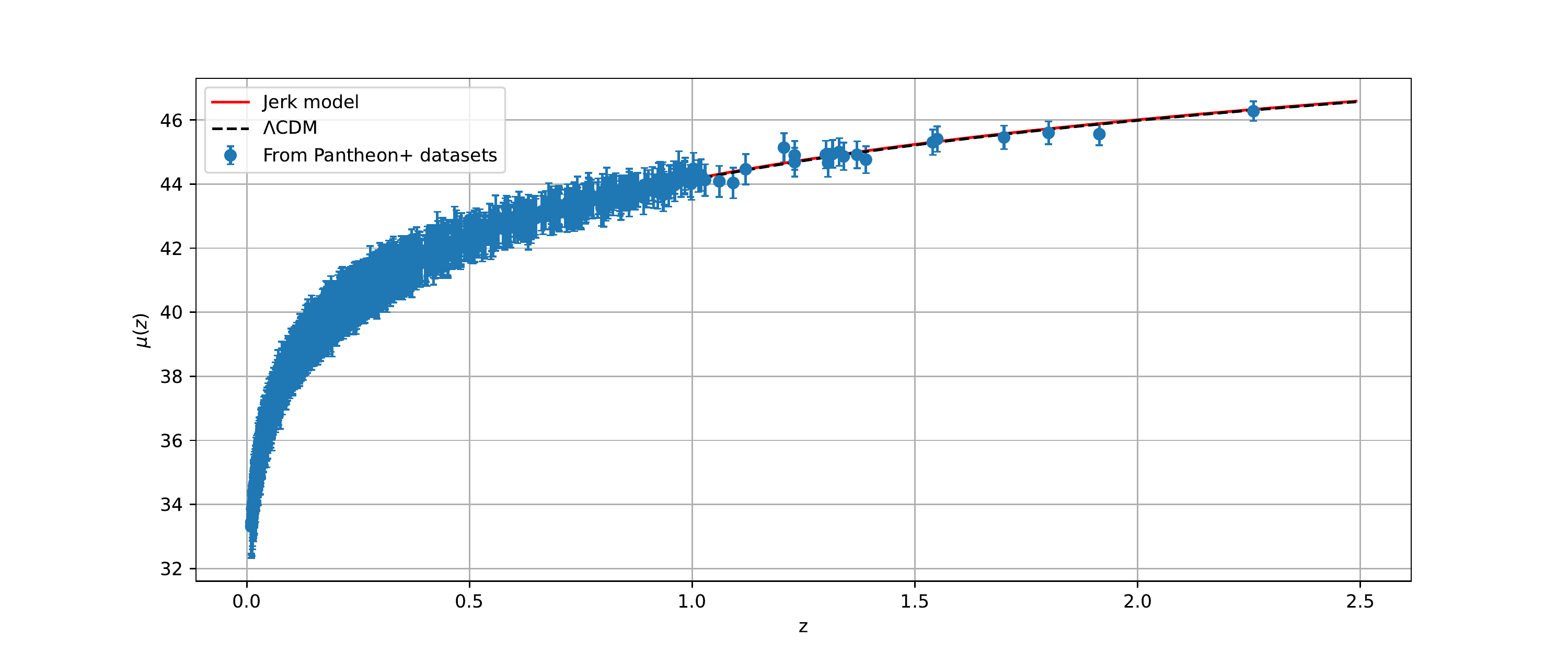}}
\caption{A good fit to the 1701 points of the Pantheon+ datasets is displayed in the plot of $\mu(z)$ versus the redshift $z$ for our jerk model, which is shown in red, and $\Lambda$CDM, which is shown in black dashed lines.}
\label{Muzcond}
\end{figure}
\end{widetext}

\begin{table}[!htp]
\centering
\begin{tabular}{c|cc}
\hline
\multicolumn{1}{c|}{Datasets} & \multicolumn{2}{c}{OHD + Pantheon+} \\
\hline
\multicolumn{1}{c|}{Model} & \multicolumn{1}{c}{$\Lambda$CDM}& \multicolumn{1}{c}{Jerk model}\\
\hline
 $\Omega_m$  &$0.282\pm 0.0118$& -
 \\[0.1cm]
  $h$  & $0.71\pm 0.0082$& -
 \\[0.1cm]
 $C$  & - & $0.3133\pm0.0095$ 
 \\[0.1cm]
 $\lambda$  & - &  $90.74\pm 1.632$
 \\[0.1cm]
  $M$ & $-19.3235\pm0.02317$ & $-19.324\pm 0.0238$
 \\[0.1cm]
\hline
 $\chi^2_{min}$  &$1564.111 $& $1564.13 $  \\[0.1cm]
\hline
\end{tabular}
\caption{Summary of the mean$\pm1\sigma$ of the cosmological parameters for the $\Lambda$CDM and jerk models, using the OHD + Pantheon+ datasets.}
\label{tab}
\end{table}

\begin{figure}[h]
\centerline{\includegraphics[scale=0.65]{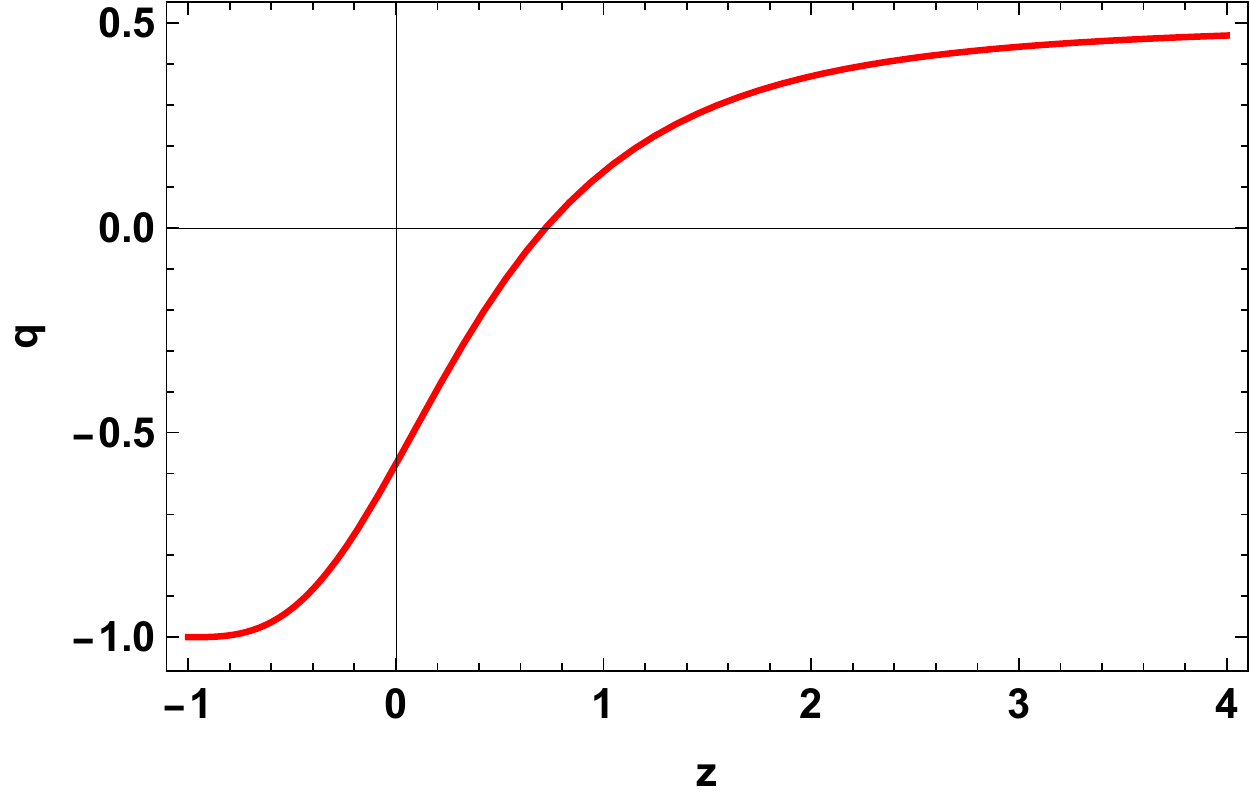}}
\caption{The graphical behavior of deceleration parameter in terms of $z$
i.e. Eq. (\protect\ref{Hz}) with the constraint values from OHD + Pantheon+
datasets.}
\label{fig2}
\end{figure}

\subsection{Energy conditions}

The energy conditions (ECs) are a set of alternative conditions that are
used to put additional constraints on the validity of the constructed
cosmological model and have many applications in theoretical cosmology. For
example these conditions play an important role in GR as they help to prove
 theorems about the presence of the singularity of space-time and black
holes \cite{Wald}. In the context of this work, ECs are exploited in order
to predict the acceleration phase of the Universe. Like these conditions can
be obtained from the famous Raychaudhury equations, which forms are \cite%
{Raychaudhuri, Nojiri, Ehlers} 
\begin{equation}
\frac{d\theta }{d\tau }=-\frac{1}{3}\theta ^{2}-\sigma _{\mu \nu }\sigma
^{\mu \nu }+\omega _{\mu \nu }\omega ^{\mu \nu }-R_{\mu \nu }u^{\mu }u^{\nu
}\,,  \label{R1}
\end{equation}%
\begin{equation}
\frac{d\theta }{d\tau }=-\frac{1}{2}\theta ^{2}-\sigma _{\mu \nu }\sigma
^{\mu \nu }+\omega _{\mu \nu }\omega ^{\mu \nu }-R_{\mu \nu }n^{\mu }n^{\nu
}\,,  \label{R2}
\end{equation}

where $n^{\mu}$, $\theta$, $\omega_{\mu\nu}$ and $\sigma ^{\mu \nu}$ are the
null vector, the expansion factor, the rotation and the shear associated
with the vector field $u^{\mu}$, respectively. In Weyl geometry with the
existence of non-metricity scalar $Q$, the Raychaudhury equations take
various forms, for more details see \cite{Arora}. Next, the above equations (%
\ref{R1}) and (\ref{R2}) fulfill the conditions 
\begin{align}
R_{\mu \nu }u^{\mu }u^{\nu }& \geq 0\,, \\
R_{\mu \nu }n^{\mu }n^{\nu }& \geq 0\,.
\end{align}

Thus, if we examine the perfect fluid distribution of cosmological matter,
the ECs for $f(Q)$ gravity are given as follows \cite{Mandal1},

\begin{itemize}
\item Weak energy conditions (WEC) if $\rho _{eff}\geq 0$, $\rho
_{eff}+p_{eff}\geq 0$.

\item Null energy condition (NEC) if $\rho _{eff}+3p_{eff}\geq 0$.

\item Dominant energy conditions (DEC) if $\rho _{eff}\geq 0$, $%
|p_{eff}|\leq \rho$.

\item Strong energy conditions (SEC) if $\rho _{eff}+3p_{eff}\geq 0$.
\end{itemize}

By taking Eqs. (\ref{eff1}) and (\ref{eff2}) in the WEC, NEC, and DEC
constraints, we can prove that

\begin{itemize}
\item Weak energy conditions (WEC) if $\rho\geq 0$, $\rho+p\geq 0$.

\item Null energy condition (NEC) if $\rho+p\geq 0$.

\item Dominant energy conditions (DEC) if $\rho \geq 0$, $|p|\leq \rho$.
\end{itemize}

These results are in concordance with those of Capozziello et al. \cite%
{Capozziello1}. In the case of the SEC condition, we find 
\begin{equation}
\rho +3\,p-6\,\dot{f}_{Q}\,H+f\geq 0\,.
\end{equation}

Now, using the above ECs, we can check the validity of our cosmological
models in the following sections.

\section{Cosmological $f\left( Q\right) $\ models}

\label{sec4}

In this section, we will discuss the proposed cosmological models and some
of their physical properties such as the energy density, pressure and
equation of state (EoS) parameter using the general solution of the jerk model parameter. In addition, we will verify our cosmological
models with the help of the ECs described in the previous section. Here, we
will propose two models of $f(Q)$ gravity. In the first model, we will
assume a linear form of $f(Q)$. Then, in the second model we will take a
non-linear functional form of $f(Q)$ gravity.

\subsection{Linear model $f\left( Q\right) =\protect\alpha Q+\protect\beta $}

In this subsection, we presume the following simplest linear form of the $%
f(Q)$ function i.e. 
\begin{equation}
f\left( Q\right) =\alpha Q+\beta  \label{f1}
\end{equation}%
where $\alpha $ and $\beta $ are free parameters. The motivation behind this
linear form is the cosmological constant, despite the problems it faces, it
is considered to be the most successful model among the alternatives offered
in cosmology. The results of this model have been discussed in several
contexts \cite{Solanki, Oleksii, Koussour1, Koussour2}.

Using Eqs. (\ref{F11}) and (\ref{H}), we get the energy density of the
Universe in the form 
\begin{equation}
\rho =\frac{3\beta \left( e^{2\lambda t}-1\right) ^{2}-8\alpha \lambda
^{2}\left( e^{2\lambda t}+1\right) ^{2}}{6\left( e^{2\lambda t}-1\right) ^{2}%
}.  \label{rho1}
\end{equation}

Again, using Eqs. (\ref{F22}) and (\ref{H}) we get the isotropic pressure of
the Universe as 
\begin{equation}
p=\frac{1}{6}\left( 8\alpha \lambda ^{2}-3\beta \right) .  \label{p1}
\end{equation}

Thus, the EoS parameter ($p=\omega \rho $) for our model is 
\begin{equation}
\omega =-\frac{\left( e^{2\lambda t}-1\right) ^{2}\left( 3\beta -8\alpha
\lambda ^{2}\right) }{3\beta \left( e^{2\lambda t}-1\right) ^{2}-8\alpha
\lambda ^{2}\left( e^{2\lambda t}+1\right) ^{2}}.  \label{EoS1}
\end{equation}

\begin{figure}[h]
\centerline{\includegraphics[scale=0.65]{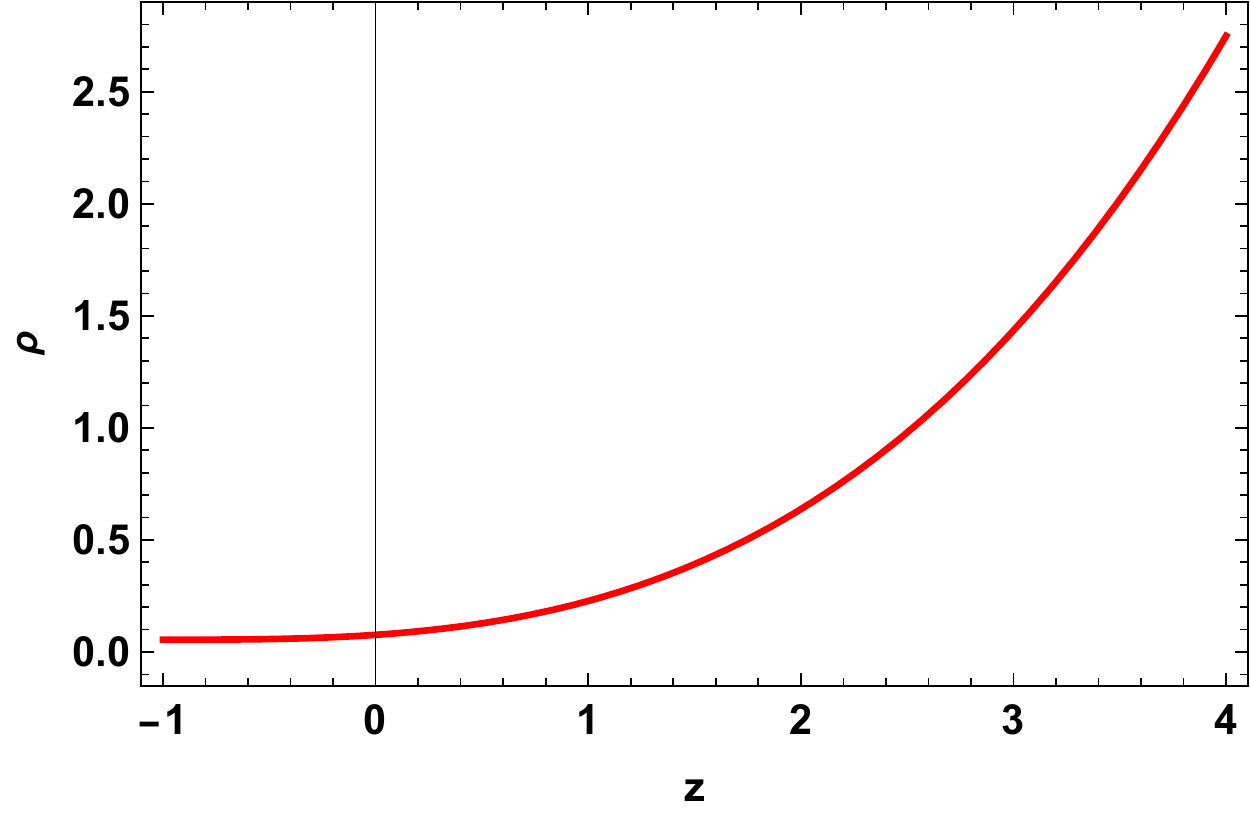}}
\caption{The graphical behavior of energy density in terms of $z$ with $%
\protect\alpha =-0.5$ and $\protect\beta =2$ for the specific case of $%
f\left( Q\right) =\protect\alpha Q+\protect\beta $.}
\label{fig3}
\end{figure}

\begin{figure}[h]
\centerline{\includegraphics[scale=0.65]{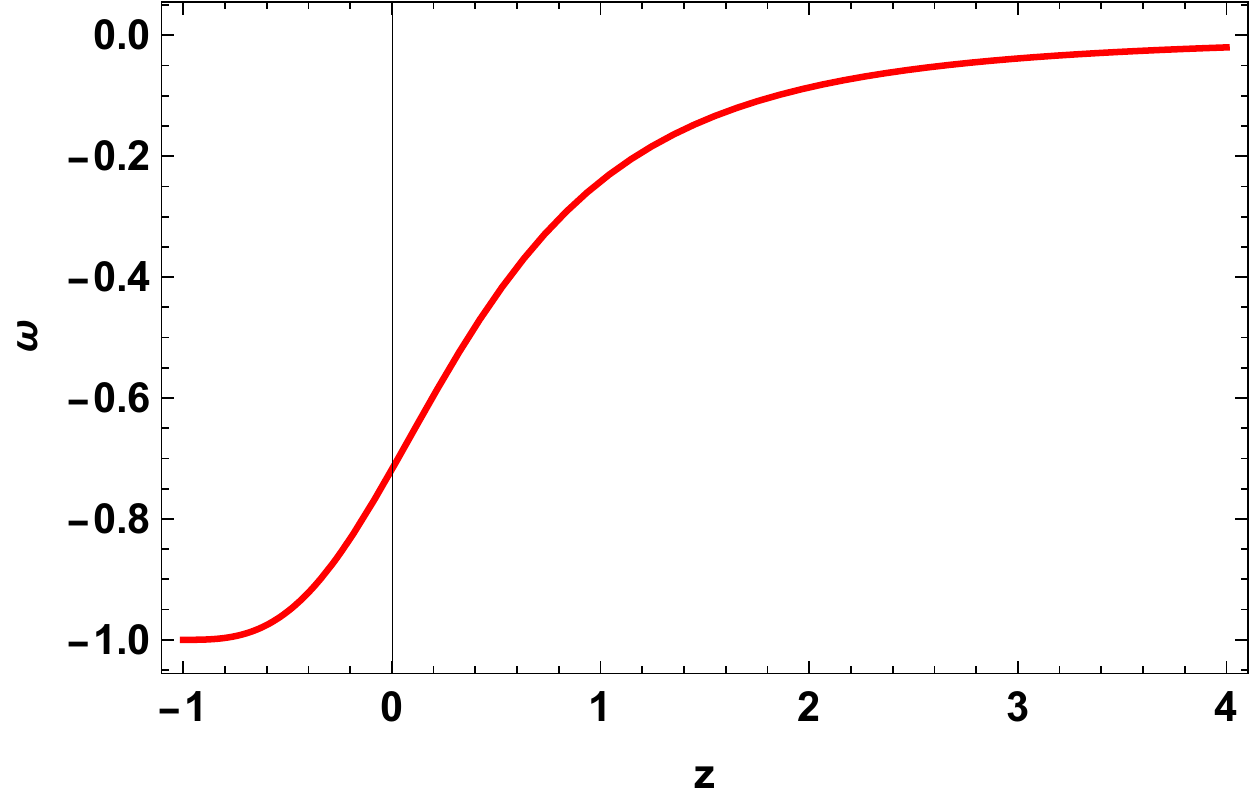}}
\caption{The graphical behavior of EoS parameter in terms of $z$ with $%
\protect\alpha =-0.5$ and $\protect\beta =2$ for the specific case of $%
f\left( Q\right) =\protect\alpha Q+\protect\beta $.}
\label{fig5}
\end{figure}

\begin{figure}[h]
\centerline{\includegraphics[scale=0.65]{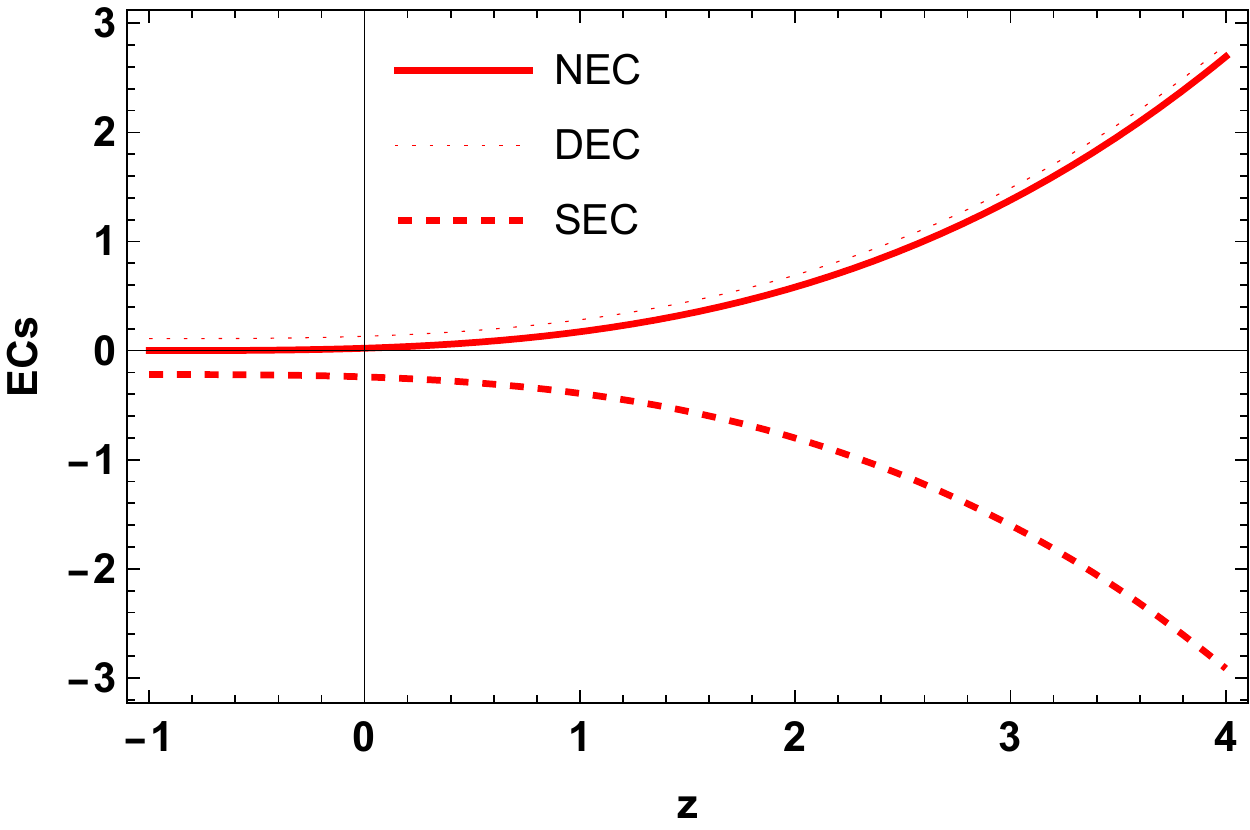}}
\caption{The graphical behavior of ECs in terms of $z$ with $\protect\alpha %
=-0.5$ and $\protect\beta =2$ for the specific case of $f\left( Q\right) =%
\protect\alpha Q+\protect\beta $.}
\label{fig6}
\end{figure}

From Fig. \ref{fig3}, we can observe that as the Universe expands, its
energy density stays positive and decreasing function of cosmic time (or
increasing function of redshift). Also, they tend to zero in the future.
Moreover, the plot for EoS parameter in Fig. \ref{fig5} shows
quintessence-like behavior in the present, converges to the $\Lambda $CDM
model in the future and to the dust matter in the past. Also, the present
value of the EoS parameter corresponding to the OHD+Pantheon+ is $\omega
_{0}=-0.7161$. Now, using Eqs. (\ref{rho1}) and (\ref{p1}) in the above ECs,
we have plotted the behavior of NEC, DEC, and SEC in terms of the cosmological
redshift in Fig. \ref{fig6}. From this figure, it can be clearly seen that
all the ECs are satisfied while the SEC is violated. This violation of the
SEC is the evidence of the validity of the proposed cosmological model, and
thus predicts the accelerating phase of the Universe.

\begin{figure}[h]
\centerline{\includegraphics[scale=0.65]{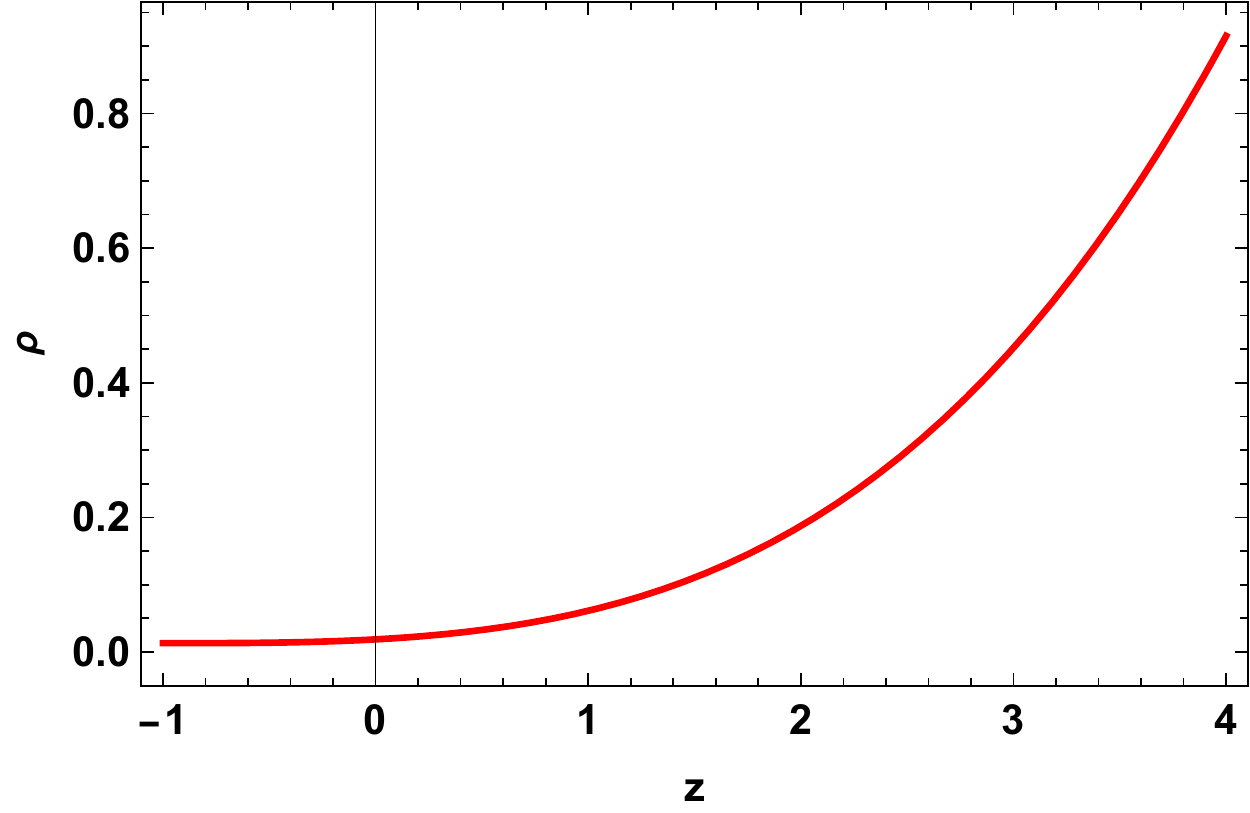}}
\caption{The graphical behavior of energy density in terms of $z$ with $m=-5$
and $n=1.08$ for the specific case of $f\left( Q\right) =Q+mQ^{n}$.}
\label{fig7}
\end{figure}

\begin{figure}[h]
\centerline{\includegraphics[scale=0.65]{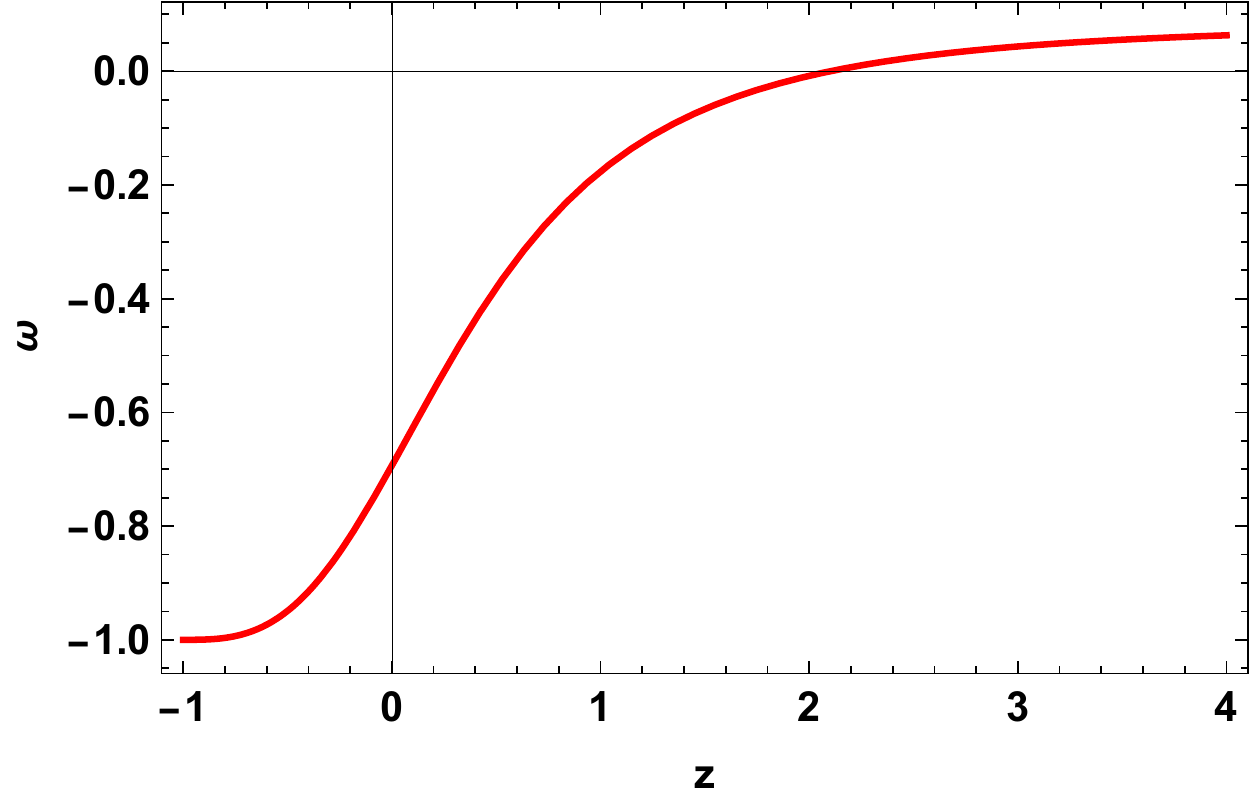}}
\caption{The graphical behavior of EoS parameter in terms of $z$ with $m=-5$
and $n=1.08$ for the specific case of $f\left( Q\right) =Q+mQ^{n}$.}
\label{fig9}
\end{figure}

\begin{figure}[h]
\centerline{\includegraphics[scale=0.65]{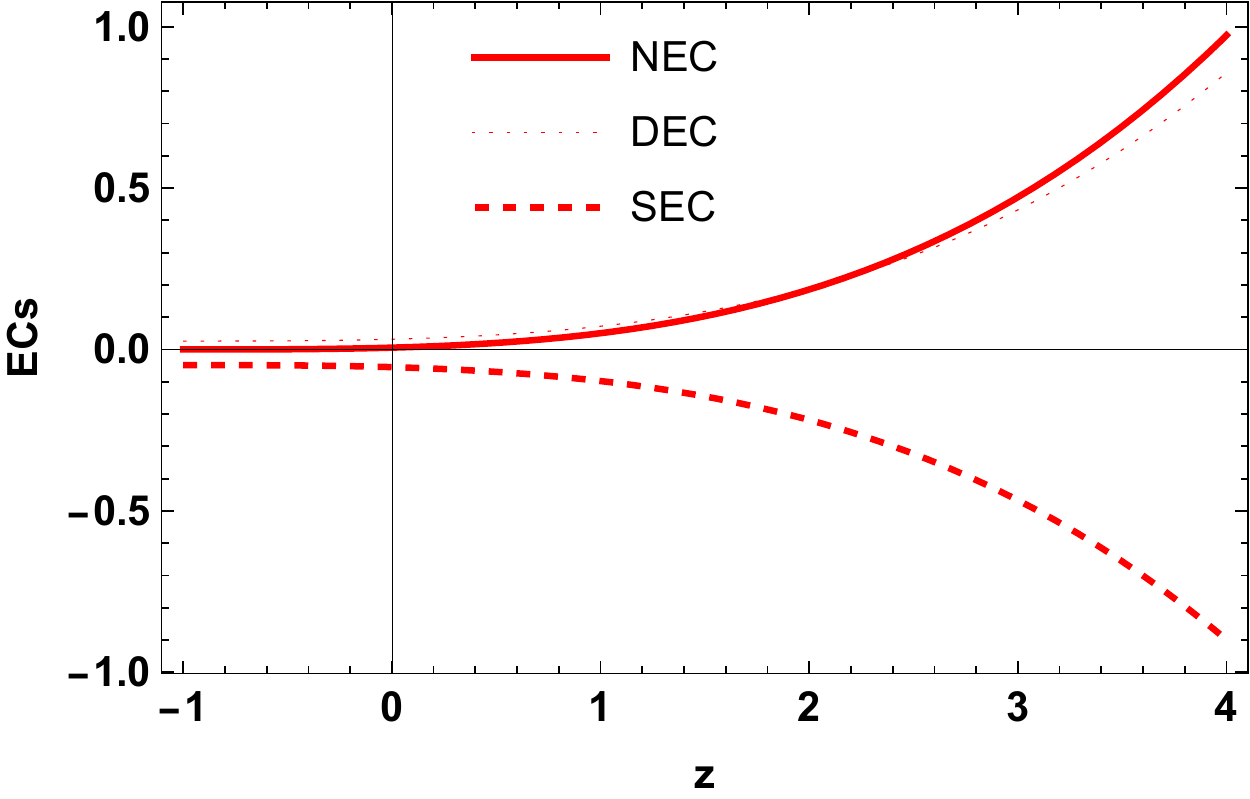}}
\caption{The graphical behavior of ECs in terms of $z$ with $m=-5$ and $%
n=1.08$ for the specific case of $f\left( Q\right) =Q+mQ^{n}$.}
\label{fig10}
\end{figure}

\subsection{Non-linear model $f\left( Q\right) =Q+mQ^{n}$}

Here, for the second model, we discuss the non-linear functional form of $%
f(Q)$, 
\begin{equation}
f\left( Q\right) =Q+mQ^{n}  \label{f2}
\end{equation}%
where $m$ and $n$ are free parameters. Also, this specific form has been
considered in many cosmological contexts \cite{Shekh, Mandal2, Oleksii}.

Thus, for this specific choice, we get the energy density, the isotropic
pressure and the EoS parameter as 
\begin{widetext} 
\begin{equation}
\rho =\frac{2^{3n-1}3^{-n-1}\left( -3m(2n-1)\left( e^{2\lambda t}-1\right)
^{2}\left( \frac{9}{4}H^{2}\right) ^{n}-\lambda ^{2}3^{n}8^{1-n}\left(
e^{2\lambda t}+1\right) ^{2}\right) }{\left( e^{2\lambda t}-1\right) ^{2}},
\label{rho2}
\end{equation}
\begin{equation}
p=\frac{3^{-n-1}\left( 3m8^{n}(2n-1)\left( (2-4n)e^{2\lambda t}+e^{4\lambda
t}+1\right) \left( \frac{9}{4}H^{2}\right) ^{n}+8\lambda ^{2}3^{n}\left(
e^{2\lambda t}+1\right) ^{2}\right) }{2\left( e^{2\lambda t}+1\right) ^{2}},
\label{p2}
\end{equation}
and 
\begin{equation}
\omega =-\frac{\left( e^{2\lambda t}-1\right) ^{2}\left( 3m8^{n}(2n-1)\left(
(2-4n)e^{2\lambda t}+e^{4\lambda t}+1\right) \left( \frac{9}{4}H^{2}\right)
^{n}+8\lambda ^{2}3^{n}\left( e^{2\lambda t}+1\right) ^{2}\right) }{\left(
e^{2\lambda t}+1\right) ^{2}\left( 3m8^{n}(2n-1)\left( e^{2\lambda
t}-1\right) ^{2}\left( \frac{9}{4}H^{2}\right) ^{n}+8\lambda ^{2}3^{n}\left(
e^{2\lambda t}+1\right) ^{2}\right) },  \label{EoS2}
\end{equation}
\end{widetext} respectively.

From Fig. \ref{fig7}, it is clear that the energy density of the Universe is
an increasing function of cosmological redshift and remains positive as the
Universe expands. In addition, it tends to zero in the future. Further, the
plot for EoS parameter in Fig. \ref{fig9} shows quintessence-like behavior
in the present and converges to the $\Lambda $CDM model in the future and to
the dust matter in the past. Further, the present value of the EoS parameter
corresponding to the OHD+Pantheon+ is $\omega _{0}=-0.7009$. For this case,
using Eqs. (\ref{rho2}) and (\ref{p2}) in the above ECs, we have plotted the
behavior of NEC, DEC, and SEC in terms of cosmological redshift in Fig. \ref%
{fig10}. From this figure, it is clear that all the ECs are satisfied but
SEC is violated. This violation of the SEC is the evidence of the validity
of the proposed cosmological model, and thus predicts the accelerating phase
of the Universe.

\section{Discussions and conclusions}

\label{sec5}

The standard model of cosmology ($\Lambda $CDM) is most widely accepted
today as it has been able to explain a large number of observed phenomena:
the expansion of the Universe, the existence of the cosmic microwave
background and the big bang nucleosynthesis. However, as we have pointed out
in the introduction, $\Lambda $CDM could not explain dark energy (DE) and
other issues \cite{Bourakadi1, Bourakadi2}. These puzzles prompted many
authors to search for suitable alternatives, and some scientists went so far
as to suggest a modification of general relativity, the theoretical basis of
the $\Lambda $CDM model.

In this paper, we have discussed one of these recently proposed theories
which has attracted the attention of many researchers i.e. $f(Q)$ gravity
where the non-metricity $Q$ is the basis of gravitational interactions with
zero curvature and torsion. We have studied a homogeneous and isotropic FLRW
space-time in the framework of this modified theory and the help of the jerk parameter. The jerk parameter can be employed in a number of scenarios. Chakrabarti et al. \cite{Chakrabarti} proposed a reconstruction of extended teleparallel $f(T)$ gravity using this parameter. In this scenario, we used the jerk parameter to study the late-time expansion of the Universe in $f(Q)$ gravity. So, we have briefly described the
mathematical formalism of $f(Q)$ gravity, then we have derived the field equations in the FLRW
space-time for the content of the Universe in the form of a perfect fluid.
Moreover, we have used $31$ data points of OHD and $1701$ data
points of Pantheon+ to constrain the model parameters. The current Hubble rate and the
deceleration parameter derived from the best fit of Markov Chain Monte Carlo (MCMC)
are in agreement with those of the Planck data \cite{Planck2018}. Moreover, we
combined OHD + Pantheon+ datasets with recently published Pantheon+ datasets to get the model parameters that fit the data the best. The results of
the best fit is $\lambda=90.74\pm 1.632$ and  $C=0.3133\pm0.0095$. In the same context, Mukherjee et al. \cite{Mukherjee2019} constrained a variables jerk parameters by means of MCMC. Furthermore, Ayuso et al. \cite{Ayuso} employed the MCMC approach to constrain the general power model of $f(Q)$ gravity and determine the best-fit of cosmological parameter.

Next, we have considered two functional forms of $f(Q)$ gravity,
specifically, a linear and a non-linear form. We have analyzed the behavior
of different cosmological parameters such as energy density, pressure and
EoS parameter for both models. we have also checked all energy conditions in
order to ensure the validity of our proposed cosmological models. For both
models, Figs. \ref{fig3} and \ref{fig7}, we have found that the energy
density of the Universe is a decreasing function of cosmic time (or
increasing function of redshift) and remains positive as the Universe
expands. Furthermore, from Figs. \ref{fig5} and \ref{fig9}, we have observed
that the EoS parameter behaves like a quintessence dark energy model in the
vicinity of our present time, while in the future and in the past it behaves
as the $\Lambda $CDM model and the dust matter, respectively for both
models. Finally, from the energy conditions as shown in Figs. \ref{fig6} and %
\ref{fig10} we can conclude that all the energy conditions are satisfied for
both models while SEC is violated. The results above demonstrate that our
proposed cosmological models are in strong agreement with today's
observations.

\textbf{Data availability} There are no new data associated with this article

\textbf{Declaration of competing interest} The authors declare that they
have no known competing financial interests or personal relationships that
could have appeared to influence the work reported in this paper.\newline

\acknowledgments 
We are very much grateful to the
honorable referee and to the editor for the illuminating suggestions that
have significantly improved our work in terms of research quality, and
presentation.

%%%%%%%%%%%%%%%%%%%%%%%%%%%%%%%%%%%%%%%%%%%%%%%%%%%%%%%%%%%%%%%%%%%%%%%%%%%
%%%%%%%%%%%%%%%%%%%%%%%%%%%%%%%%%%%%%%%%%%%%%%%%%%%%%%%%%%%%%%%%%%%%%%%%%%%


\begin{thebibliography}{0}%
\makeatletter
\providecommand \@ifxundefined [1]{%
 \@ifx{#1\undefined}
}%
\providecommand \@ifnum [1]{%
 \ifnum #1\expandafter \@firstoftwo
 \else \expandafter \@secondoftwo
 \fi
}%
\providecommand \@ifx [1]{%
 \ifx #1\expandafter \@firstoftwo
 \else \expandafter \@secondoftwo
 \fi
}%
\providecommand \natexlab [1]{#1}%
\providecommand \enquote  [1]{``#1''}%
\providecommand \bibnamefont  [1]{#1}%
\providecommand \bibfnamefont [1]{#1}%
\providecommand \citenamefont [1]{#1}%
\providecommand \href@noop [0]{\@secondoftwo}%
\providecommand \href [0]{\begingroup \@sanitize@url \@href}%
\providecommand \@href[1]{\@@startlink{#1}\@@href}%
\providecommand \@@href[1]{\endgroup#1\@@endlink}%
\providecommand \@sanitize@url [0]{\catcode `\\12\catcode `\$12\catcode
  `\&12\catcode `\#12\catcode `\^12\catcode `\_12\catcode `\%12\relax}%
\providecommand \@@startlink[1]{}%
\providecommand \@@endlink[0]{}%
\providecommand \url  [0]{\begingroup\@sanitize@url \@url }%
\providecommand \@url [1]{\endgroup\@href {#1}{\urlprefix }}%
\providecommand \urlprefix  [0]{URL }%
\providecommand \Eprint [0]{\href }%
\providecommand \doibase [0]{https://doi.org/}%
\providecommand \selectlanguage [0]{\@gobble}%
\providecommand \bibinfo  [0]{\@secondoftwo}%
\providecommand \bibfield  [0]{\@secondoftwo}%
\providecommand \translation [1]{[#1]}%
\providecommand \BibitemOpen [0]{}%
\providecommand \bibitemStop [0]{}%
\providecommand \bibitemNoStop [0]{.\EOS\space}%
\providecommand \EOS [0]{\spacefactor3000\relax}%
\providecommand \BibitemShut  [1]{\csname bibitem#1\endcsname}%
\let\auto@bib@innerbib\@empty
%</preamble>
\end{thebibliography}%


\begin{thebibliography}{99}
\bibitem{SN1} A.G. Riess et al., \textit{Astron. J}. \textbf{116}, 1009
(1998).

\bibitem{SN2} S. Perlmutter et al., \textit{Astrophys. J}. \textbf{517}, 565
(1999).

\bibitem{CMB1} R.R. Caldwell, M. Doran, \textit{Phys. Rev.} D 69, 103517
(2004).

\bibitem{CMB2} Z.Y. Huang et al., \textit{JCAP} \textbf{0605}, 013 (2006).

\bibitem{LS1} T. Koivisto, D.F. Mota, \textit{Phys. Rev. D} \textbf{73},
083502 (2006).

\bibitem{LS2} S.F. Daniel, \textit{Phys. Rev. D} \textbf{77}, 103513 (2008).

\bibitem{BAO1} D.J. Eisenstein et al., \textit{Astrophys. J}. \textbf{633},
560 (2005).

\bibitem{BAO2} W.J. Percival at el., \textit{Mon. Not. R. Astron. Soc}. 
\textbf{401}, 2148 (2010).

\bibitem{WMAP1} C.L. Bennett et al., \textit{Astrophys. J. Suppl}. \textbf{%
148}, 119-134 (2003).

\bibitem{WMAP2} D.N. Spergel et al., [WMAP Collaboration], \textit{%
Astrophys. J. Suppl.} \textbf{148}, 175 (2003).

\bibitem{WMAP9} G. Hinshaw et al., \textit{Astrophys. J. Suppl}. \textbf{208}%
, 19 (2013).

\bibitem{Planck2018} N. Aghanim et al., \textit{Astron. Astrophys.} \textbf{%
641}, A6 (2020).

\bibitem{Weinberg} S.Weinberg, \textit{Rev. Mod. Phys.} \textbf{61}, 1
(1989).

\bibitem{quintessence} B. Ratra and P.J.E. Peebles, \textit{Phys. Rev. D} 
\textbf{37}, 3406 (1998).

\bibitem{phantom} M. Sami and A. Toporensky, \textit{Mod. Phys. Lett. A} 
\textbf{19}, 1509 (2004).

\bibitem{kessence} C. Armendariz-Picon et al., \textit{Phys. Rev. Lett.} 
\textbf{85}, 4438 (2000).

\bibitem{Chameleon} J. Khoury and A. Weltman, \textit{Phys. Rev. Lett.} 
\textbf{93}, 171104 (2004).

\bibitem{tachyon} T. Padmanabhan, \textit{Phys. Rev. D} \textbf{66}, 021301
(2002).

\bibitem{Cgas1} M. C. Bento et al., \textit{Phys. Rev. D} \textbf{66},
043507 (2002).

\bibitem{Cgas2} R. Zarrouki and M. Bennai, \textit{Phys. Rev. D} \textbf{82,}
123506 (2010).

\bibitem{ouali2015} M. Bouhmadi-L\'opez, A. Errahmani, P. Martín-Moruno, T.
Ouali, Y. Tavakoli, \textit{Internat. J. Modern Phys. D} \textbf{24 (10),}
1550078 (2015).

\bibitem{bouhmadi2017} J. Morais, M. Bouhmadi-L\'opez, K. Sravan Kumar, J.
Marto, Y. Tavakoli, \textit{Phys. Dark Univ.} \textbf{15,} 7 (2017).

\bibitem{bouhmadi2018} M. Bouhmadi-L\'opez, D. Brizuela, I. Garay, \textit{%
J. Cosmol. Astropart. Phys.} \textbf{1809 (09),} 031 (2018).

\bibitem{ouali2019} A. Bouali, I. Albarran, M. Bouhmadi-L\'opez, T. Ouali, 
\textit{Phys. Dark Univ.} \textbf{26,} 100391 (2019).

\bibitem{ouali2021} A. Bouali, I. Albarran, M. Bouhmadi-L\'opez, A.
Errahmani, T. Ouali, \textit{Phys. Dark Univ.} \textbf{34,} 100907(2021).

%\bibitem{fR} S. Capozziello et al., \textit{Phys. Rev. D} \textbf{76,}
%104019 (2007).

%\bibitem{fT} M. Koussour and M. Bennai, \textit{Class. Quantum Gravity} 
%\textbf{39}, 105001 (2022).

%\bibitem{fG} M. Koussour et al., \textit{Nucl. Phys. B.} \textbf{978},
%115738 (2022).

\bibitem{Capozziello} S. Capozziello et al., \textit{Phys. Lett. B} \textbf{%
632}, 597 (2006).

\bibitem{Blandford} R.D. Blandford et al., arXiv preprint
arXiv:astro-ph/0408279 (2004).

\bibitem{Chiba} T. Chiba and T. Nakamura, \textit{Prog. Theor. Phys.} 
\textbf{100}, 1077 (1998).

\bibitem{Sahin} V. Sahin, arXiv preprint arXiv:astro-ph/0211084 (2002).

\bibitem{Visser1} M. Visser, \textit{Class. Quantum Gravity} \textbf{21},
2603 (2004).

\bibitem{Visser2} M. Visser, \textit{Gen. Relativ. Gravit.} \textbf{37},
1541 (2005).

\bibitem{Alam} U. Alam et al., \textit{Mon. Not. Roy. Astron. Soc.} \textbf{%
344}, 1057--1074 (2003).

\bibitem{Rapetti} R.D. Rapetti et al., \textit{Mon. Not. Roy. Astron. Soc}. 
\textbf{375}, 1510 (2007).

\bibitem{Zubair} M. Zubair and L. R. Durrani, \textit{Eur. Phys. J. Plus} 
\textbf{135}, 8 (2020).

\bibitem{Lu} J. Lu, L. Xu and M. Liu, \textit{Phys. Lett. B} 
\textbf{699}, 8 (2011).

\bibitem{Xu} Y. Xu et al., \textit{Eur. Phys. J. C} \textbf{79}, 8 (2019).

\bibitem{Mandal1} S. Mandal et al., \textit{Phys. Rev. D} \textbf{102},
024057 (2020).

\bibitem{Mandal2} S. Mandal et al., \textit{Phys. Rev. D }\textbf{102},
124029 (2020).

\bibitem{Harko1} T. Harko et al., \textit{Phys. Rev. D} \textbf{98}, 084043
(2018).

\bibitem{Dimakis} N. Dimakis et al., \textit{Class. Quantum Grav}. \textbf{38%
}, 225003 (2021).

\bibitem{Shekh} S. H. Shekh, \textit{Phys. Dark Universe} \textbf{33},
100850 (2021).

\bibitem{Koussour1} M. Koussour et al., \textit{J. High Energy Astrophys 
\textbf{35}, }43-51 (2022).

\bibitem{Koussour2} M. Koussour et al., \textit{Phys. Dark Universe} \textbf{%
36}, 101051 (2022).

\bibitem{Jimenez1} J. B. Jimenez et al., \textit{Phys. Rev. D} \textbf{98},
044048 (2018).

\bibitem{Jimenez2} J. B. Jimenez et al., \textit{Phys. Rev. D} \textbf{101},
103507 (2020).

\bibitem{para1} J. V. Cunha and J. A. S. D. Lima, \textit{Mon. Notices Royal Astron. Soc.} \textbf{390},
1 (2008).

\bibitem{para2} E. Mortsell and C. Clarkson, \textit{J. Cosm. Astropar. Phys.} \textbf{2009},
01 (2009).

\bibitem{para3} S. K. J. Pacif et al., \textit{Int. J. Geom. Meth. Mod. Phys.} \textbf{14},
7 (2017).

\bibitem{Lazkoz} R. Lazkoz et al.,\textit{\ Phys. Rev. D} \textbf{100},
104027 (2019).

\bibitem{Chakrabarti} Chakrabarti et al., \textit{Eur. Phys. J. C} \textbf{79%
}, 454 (2019).

\bibitem{MCMC} L. E. Padilla et al., \textit{Universe} \textbf{97} (2021)
213. [{arXiv:1903.11127}]

\bibitem{Sharov} G. S. Sharov V. O. Vasiliev, \textit{Appl. Math. Model} 
\textbf{6} (2018) 1. [{arXiv:1807.07323}]

\bibitem{data3} S. Joan , \textit{Phys. Rev. D }. \textbf{71} (2005)
255-262. [{arXiv:1209.0210}]

\bibitem{data2} J. E. Bautista et al., \textit{Astron. Astrophys.} \textbf{%
603} (2017) A12. [{arXiv:1702.00176}]

\bibitem{data4} D. Brout et al., \textit{Astrophys . J.} \textbf{938}
(2022) 110. {[arXiv:2202.04077v2]}



\bibitem{Wald} R. M. Wald, \textit{Chicago, IL: University of Chicago Press}%
, (1984).

\bibitem{Raychaudhuri} A. Raychaudhuri, \textit{Phys. Rev. D} \textbf{98},
1123 (1955).

\bibitem{Nojiri} S. Nojiri and S. D. Odintsov, \textit{Int. J. Geom. Methods
Mod. Phys}. \textbf{04}, 115 (2007).

\bibitem{Ehlers} J. Ehlers, \textit{Int. J. Mod. Phys. D} \textbf{15}, 1573
(2006).

\bibitem{Arora} S. Arora, et al., \textit{Phys. Dark Universe} \textbf{31},
100790 (2021).

\bibitem{Capozziello1} S. Capozziello, S. Nojiri, and S. D. Odintsov, 
\textit{Phys. Lett. B} \textbf{781}, 99 (2018).

\bibitem{Solanki} R. Solanki et al., arXiv preprint arXiv:2201.06521 (2022).

\bibitem{Oleksii} Sokoliuk et al., arXiv preprint arXiv:2201.00743 (2022).

\bibitem{Bourakadi1} K. El Bourakadi et al., \textit{Eur. Phys. J. Plus} 
\textbf{136}, 8 (2021).

\bibitem{Bourakadi2} K. El Bourakadi et al., \textit{Eur. Phys. J. C} 
\textbf{81}, 12 (2021).


\bibitem{Mukherjee2019} A. Mukherjee and N. Banerjee., \textit{Phys. Rev. D} 
\textbf{93}, 043002 (2016).

\bibitem{Ayuso} I. Ayuso et al., \textit{Phys. Rev. D} 
\textbf{103}, 6 (2021).

\end{thebibliography}
\end{document}